# Zigzag antiferromagnetic ground state with anisotropic correlation lengths in the quasi-2D honeycomb lattice compound $Na_2Co_2TeO_6$

A. K. Bera, S. M. Yusuf,[*] and Amit Kumar

*Solid State Physics Division, Bhabha Atomic Research Centre, Mumbai 400085, India*

C. Ritter

*Institut Laue-Langevin, 71 Avenue des Martyrs, 38000 Grenoble, France*

(Dated: June 9, 2016)

## Abstract

The crystal structure, magnetic ground state, and the temperature dependent microscopic spin-spin correlations of the new frustrated honeycomb lattice antiferromagnet $Na_2Co_2TeO_6$ have been investigated by powder neutron diffraction. A three dimensional (3D) long-range antiferromagnetic (AFM) ordering has been found below $T_N \sim 24.8$ K. The magnetic ground state, determined to be zigzag antiferromagnet and characterized by a propagation vector **k** = (1/2 0 0), occurs due to the competing exchange interactions up to third nearest neighbors within the honeycomb lattice. The exceptional existence of a limited magnetic correlation length along the *c* axis (perpendicular to the honeycomb layers in the *ab* planes) has been found within the 3D long-range ordered state, even at 1.8 K, well below $T_N \sim 24.8$ K. The observed limited correlation along the *c* axis is explained by the disorder distribution of the Na ions within the intermediate layers between honeycomb planes. The reduced ordered moments $m_{Co(1)}$ = 2.77(3) $\mu_B/Co^{2+}$ and $m_{Co(2)}$ = 2.45(2) $\mu_B/Co^{2+}$ at 1.8 K reflect the persistence of spin fluctuations in the ordered state. Above $T_N \sim 24.8$ K, the presence of short-range magnetic correlations, manifested by broad diffuse magnetic peaks in the diffraction patterns, has been found. Reverse Monte Carlo analysis of the experimental diffuse magnetic scattering data reveals that the spin correlations are mainly confined within the 2D honeycomb layers (*ab* plane) with a correlation length of ~ 12 Å at 25 K. The nature of spin arrangements is found to be similar in both the short-range and long-range ordered magnetic states. This implies that the short-range correlation grows with decreasing temperature and leads to the zigzag AFM ordering at $T \leq T_N$. The present study provides a comprehensive picture of the magnetic correlations over the temperature range above and below $T_N$ and their relation to the crystal structure. The role of intermediate soft Na-layers on the magnetic coupling between honeycomb planes is discussed.

## I.     INTRODUCTION

The investigation of novel quantum states, tailored by strong quantum fluctuations and/or strong frustration, in low dimensional spin systems is an active research field in recent years. In this respect, the two dimensional (2D) honeycomb lattice model is of special interest, as it has the lowest coordination number ($z$ = 3) in two dimensions, hence, strongest possible quantum fluctuations. Unlike 2D triangular and Kagomé lattices, the honeycomb lattice with only nearest neighbor exchange interaction ($J_1$) does not show frustration and has a Néel antiferromagnetic (AFM) ground state in the classical limit ($S \rightarrow \infty$). However, frustration can be easily introduced in a honeycomb lattice by inclusion of competing next-nearest-neighbor (NNN) ($J_2$) and/or next-next-nearest-neighbor (NNNN) ($J_3$) exchange interactions. This frustrated model with $J_1$, $J_2$, and $J_3$ has been known to possess a massive degeneracy of the magnetic ground state, which, however, might be lifted either due to quantum or thermal fluctuations, the effect known as "order-by-disorder", leading to exotic ordered magnetic states and a complex magnetic phase diagram. Depending on the signs and ratios of the exchange interactions ($J_2/J_1$ and $J_3/J_1$) and the spin values, different types of ordered and quantum disordered magnetic phases are theoretically proposed for the honeycomb lattices. These include Néel, zigzag, stripy, and spiral/helical ordered states, as well as disordered quantum spin liquid and quantum paramagnetic (plaquette valencebond state) states [1–8]. Moreover, the presence of an interplanar exchange coupling between honeycomb layers can qualitatively change the microscopic nature of the magnetic ground states [9]. Furthermore, the honeycomb lattice spin systems show diverse phenomena, such as topological phase transitions (also known as Kosterlitz-Thouless transition) [10], superconductivity [11], and gapless quantum spin liquid [12]. A possible realization of the highly frustrated Kitaev-Heisenberg model has also been proposed for the honeycomb lattice [13].

Experimental efforts have been undertaken to explore and understand the unique properties of frustrated honeycomb lattice compounds. The family of compounds Ba$M_2$($X$O$_4$)$_2$ with $M$ = Co and Ni, and $X$ = P and As forms weakly coupled frustrated honeycomb lattices of magnetic ions $M$ with spin $S$ = 1/2 for Co, and $S$ = 1 for Ni. In BaCo$_2$(AsO$_4$)$_2$, the cobalt moments order abruptly at $T_c$ = 5.4 K with a helical magnetic structure [14], whereas, the isostructural compounds BaNi$_2$(PO$_4$)$_2$ and BaNi$_2$(AsO$_4$)$_2$ show collinear AFM structures (stripy and zigzag AFM structures, respectively) below 24.05 and 18.65 K, respectively, [15]. The honeycomb lattice delafossite compounds Cu$_3$Ni$_2$SbO$_6$ and Cu$_3$Co$_2$SbO$_6$ show zigzag AFM ordering in the honeycomb plane [16]. On the other hand, in the $S$ = 3/2 honeycomb Heisenberg compound Bi$_3$Mn$_4$O$_{12}$(NO$_3$) no long-range magnetic order is present due to the competing interactions between the $J_1$ and $J_2$ [12, 17, 18]. Moreover, a commensurate Néel AFM long-range order state can be induced in Bi$_3$Mn$_4$O$_{12}$(NO$_3$) by the application of a magnetic field [12, 19]. Another honeycomb compound with $S$ = 1/2, InCu$_{2/3}$V$_{1/3}$O$_3$ shows only short-range antiferromagnetic order within the 2D plane without any magnetic correlation between such planes along the $c$ axis [20].

The newly discovered families of layered honeycomb lattice compounds with general formula $A^+_2 M^{2+}_2 Te^{6+}O_6$ (P2 type crystal structure) and $A^+_3 M^{2+}_2 X^{5+}O_6$ ($A$ = Li and Na; $X$ = Bi and Sb; and $M$ is a transition metal) (O3 type crystal structure) [16, 21–28] offer the possibility to study the role of signs and ratios of exchange interactions ($J_2/J_1$ and $J_3/J_1$), as well as of the spin value on the magnetic ground state and the magnetic properties. In these compounds, the honeycomb layers are formed by edge sharing $MO_6$ octahedra with $TeO_6$ or $XO_6$ at the center of the honeycomb lattice. For magnetic ions $M$ the network created by the edge sharing $MO_6$ octahedra provides higher order exchange interactions beyond the nearest neighbors within the honeycomb layers. The magnetic honeycomb layers are separated by the nonmagnetic layers of Na/Li. In these crystal structures, the Na and Li atoms are very diffusive leading to the compounds to be suitable as cathode materials in heavy ion rechargeable batteries. At the same time, the crystal structures are soft due to the intermediate Na/Li layers, and provide an easier and effective control of the interlayer magnetic couplings, hence, a tuning of magnetic lattice dimensionality. Among the experimentally studied compounds from the above series, a spin-gap behavior was found for the $Cu^{2+}$ ($S = 1/2$) based compounds $Na_2Cu_2TeO_6$ [22] and $Na_3Cu_2SbO_6$ [23]. Absence of long-range magnetic ordering was reported for the structurally similar compounds $Na_3LiFeSbO_6$, $Na_4FeSbO_6$, and $Li_4MnSbO_6$ [29, 30], and was related to disorder and frustration effects. Recently, density functional theory (DFT) calculations proposed a long-range zigzag AFM ordering for the compounds $Li_3Ni_2SbO_6$ and $Na_3Ni_2SbO$. An AFM ordering at low temperatures was also reported for the O3 type honeycomb compounds $Na_3M_2SbO_6$ ($M$ = Cu, Ni, and Co) [16, 31, 32], $A_3Ni_2BiO_6$ with $A$ = Na and Li [25, 33] and $Li_3Ni_2SbO_6$ [24], as well as for the P2 type compounds $Na_2M_2TeO_6$ ($M$ = Co and Ni) [31, 32, 34], however, their real quantum ground states remain unknown and await for an experimental investigation. The influence of the intermediate soft Na/Li layers on the magnetic correlations remains as well unexplored in these systems.

In view of this, the bilayer honeycomb lattice compound $Na_2Co_2TeO_6$ is of present interest. Only the crystal structure and very basic bulk magnetic properties were reported for this compound in the literature [32, 34], which reveal that $Na_2Co_2TeO_6$ orders antiferromagnetically at low temperatures without giving any details of the microscopic nature of the magnetic ground state. Here, we report the microscopic nature of the magnetic ground state, and a detailed temperature evolution of the magnetic correlations of $Na_2Co_2TeO_6$, both studied by neutron powder diffraction. The magnetic ground state is determined to be a zigzag antiferromagnet having - even deep into the 3D long-range ordered state (down to 1.8 K) - a restricted correlation length perpendicular to the honeycomb layers. The broken magnetic connections between the honeycomb layers are explained on the basis of the Na-disorder in the interconnecting soft Na layers. The analysis by the reverse Monte Carlo (RMC) method of the diffuse magnetic scattering, observed above $T_N$, reveals a 2D short-range ordering with ferromagnetic (FM) NN, AFM NNN, and AFM NNNN correlations within the honeycomb planes. The present study provides a thorough characterisation of magnetic correlations above and below the $T_N$, and their correlation to the crystal structure.

## II.    EXPERIMENTAL DETAILS

Polycrystalline samples of $Na_2Co_2TeO_6$ were synthesized by solid state reaction. A stoichiometric mixture of $Na_2CO_3$ (99.9 %), $Co_3O_4$ (99.99 %) and $TeO_2$ (99.99 %) was heated at 850 °C in air for a total period of 94 hrs with several intermediate grindings.

The powder x-ray diffraction pattern was recorded using Cu $K_\alpha$ radiation at room temperature. The ac susceptibility measurements were carried out using a commercial magnetometer (Cryogenic Co. Ltd., UK) under an ac field amplitude of 5 Oe, and a frequency of 987 Hz. Temperature dependent heat capacity was measured by an AC calorimeter (Cryogenic). Isothermal magnetizations were measured by a vibrating sample magnetometer (Cryogenic). A room temperature neutron diffraction pattern was measured using the neutron powder diffractometer-II ($\lambda$ = 1.2443 Å) at Dhruva reactor, Bhabha Atomic Research Centre, Mumbai, India. Low temperature neutron diffraction measurements were performed using the D1B ($\lambda$ = 2.524 Å) and D20 ($\lambda$ = 2.41 Å) powder diffractometers at the Institut Laue-Langevin (ILL), Grenoble, France. For these neutron diffraction measurements, powder samples were filled in a cylindrical vanadium sample container and a standard orange cryostat was used for temperature variation. The measured diffraction patterns were analyzed by using the Rietveld refinement technique (by employing the FULLPROF computer program [35]). The short-range spin-spin correlations were investigated by the reverse Monte Carlo (RMC) method based program SPINVERT [36].

## III. RESULTS AND DISCUSSIONS

### A. Crystal structure

The crystal structure of $Na_2Co_2TeO_6$ is investigated by combined x-ray and neutron diffraction at room temperature (Fig. 1). The Rietveld analysis reveals that the compound crystallizes in the hexagonal symmetry with space group $P6_322$ (No. 182). The refined values of lattice parameters, atomic positions, and isotropic thermal parameters are given in Table I. The compound $Na_2Co_2TeO_6$ displays a primitive, two-layer hexagonal crystal structure [Fig. 2(a)]. The layers consist of edge sharing $CoO_6$ and $TeO_6$ octahedra within the $ab$ planes. The edge sharing $CoO_6$ octahedra form regular honeycomb lattices, with $TeO_6$ octahedra being at the centre of the honeycomb lattices [Fig.2(b)]. Possible exchange paths for the propagation of the higher order exchange interactions ($J_2$ and/or $J_3$) beyond NN ($J_1$) on this honeycomb lattice are indicated in Figs. 2 (b-c). The details of the possible superexchange interaction pathways for $J_1$, $J_2$ and $J_3$ are given in Table II. The honeycomb planes are separated along the $c$ axis by an intermediate layer of Na ions. The Na ions are situated within the $NaO_6$ triangular prisms which connect the honeycomb layers.

In the present crystal structure of $Na_2Co_2TeO_6$, there are two types of symmetry independent Co sites, one type of Te site, one type of O site, and three independent Na sites. All the Co, Te and O sites are fully occupied. However, the three Na sites are only partially occupied. They contain different percentages of Na-ions [Na(1): ~ 67 %, Na(2): ~ 25 %, Na(3): ~ 8 %]. This leads to a highly disordered distribution of Na ions within the intermediate layers between the honeycomb layers. Moreover, the three Na triangular prismatic sites connect two adjacent honeycomb layers in different ways [Fig. 2 (d)-(f)]. Na(1) triangular prisms share edges with the $CoO_6$ and $TeO_6$ octahedra in the layers above and below, whereas Na(2) triangular prisms share faces with one Co and one Te octahedron in the layers above and below, respectively. On the other hand, Na(3) triangular prisms share faces with two Co octahedra which are situated in the layers above and below, respectively. Therefore, three interlayer couplings with different strengths are possible via the three Na sites. These special structural connections via the intermediate Na layers along with disordered distributions of Na ions play a vital role on the formation of magnetic correlations between the magnetic honeycomb layers along the *c* axis, as found in our low temperature neutron diffraction study (discussed later in the magnetic ground state and short-range magnetic correlations sections).

For both magnetic Co(1) and Co(2) sites, the $CoO_6$ octahedra are formed by six equal bond lengths [2.094(1) Å for Co(1)–O, and 2.119(1) Å for Co(2)–O; (Table III)]. However, the octahedra are found to be flattened along the *c* axis (perpendicular to the honeycomb layers) [Fig. 2(g)]. Here, all the six O-Co-O bond angles parallel to the basal plane [96.9(1)° for O–Co(1)–O, and 97.4(2)° for O–Co(2)–O; (Table III)] are larger than 90°, whereas, the bond angles across the shared edges (perpendicular to the basal plane) [78.4(1)° and 88.1(1)° for O–Co(1)–O, and 78.5(1)° and 86.82(1)° for O–Co(2)–O; (Table III)] are lower than 90°. For a regular octahedron, all the angles have a value of 90°. This octahedral distortion indicates the presence of a trigonal crystal field at the Co sites.

B.  **Bulk physical properties**

Figure 3 shows the temperature dependent ac susceptibility $\chi(T)$ and heat capacity $C_p(T)$ curves. Both the $\chi(T)$ and $C_p(T)$ curves show anomalies at ∼ 24.8 K indicating the onset of the long-range magnetic ordering. The nature of the anomalies i.e., a sharp peak in the $\chi(T)$ curve, and a $\lambda$ like peak in the $C_p(T)$ curve suggests an antiferromagnetic type magnetic ordering. The inset of Fig. 3 shows the $\chi T$ vs $T$ curve. At high temperatures above ∼ 150 K, the temperature independent constant value of $\chi T$ indicates a paramagnetic state. With decreasing temperature below ∼ 150 K, the $\chi T$ curve first deviates from the paramagnetic behavior, and then shows a sharp decrease below ∼ 50 K (well above the $T_N$ ∼ 24.8 K), where an onset of short-range magnetic correlations is found in the neutron diffraction study (discussed later).

Figure 4 shows the isothermal magnetization curves at 2 and 30 K. At 30 K (above $T_N$ ∼ 24.8 K), the $M(H)$ curve shows a linear type behaviour, whereas, at 2 K, (in the ordered AFM state; $T<T_N$ ∼ 24.8 K), the $M(H)$ curve shows an upturn at ∼ 5 T which is further confirmed by its derivative (inset of Fig. 4). A similar $M$ vs $H$ behavior for $Na_2Co_2TeO_6$ was reported earlier by Viciu $et.$ $al.$ [32]. The upturn in the $M(H)$ curve suggests a field induced magnetic transition. Such a field induced transition was reported for several honeycomb antiferromagnets [16, 33, 37] having zigzag AFM ground states as found for the present compound $Na_2Co_2TeO_6$ (discussed below). In $\alpha$-RuCl$_3$ [37], the transition was reported to be an order-disorder transition from the zigzag ordered state to a field induced paramagnetic state. A detailed study of the field induced transition in $Na_2Co_2TeO_6$ is beyond the scope of the present work.

C.  **Magnetic ground state**

The nature of the magnetic ground state has been investigated by low temperature neutron powder diffraction. The neutron diffraction patterns at 30 K (paramagnetic state) and 1.8 K (magnetically ordered state) are shown in Fig. 5. The experimental pattern at 30 K can be refined with the hexagonal crystal structure of $Na_2Co_2TeO_6$ as found at room temperature (Fig. 1). The presence of few additional weak Bragg peaks from an undetermined secondary phase is also evident. The additional peaks are also evident in the room temperature pattern (Fig. 1). Nevertheless, these additional peaks do not prevent us from determining the magnetic structure from the analysis of low temperature neutron diffraction patterns as they remain unchanged with the variation of temperature from 1.8 to 30 K. At 1.8 K, the appearance of a set of additional peaks [marked by stars in Fig. 5(b)] confirms an antiferromagnetic ground state. In order to determine the magnetic structure and the magnetic moments (without any influence from impurity peaks) we use the difference pattern between the patterns at 1.8 and 30 K.

All the magnetic peaks at 1.8 K could be indexed with a propagation vector **k** = (1/2 0 0) with respect to the hexagonal unit cell of the nuclear phase of $Na_2Co_2TeO_6$. To determine the magnetic structure compatible with the space-group symmetry, we carried out representational analysis [38–44] using the program BASIREPS from the FULLPROF package [45, 46]. The symmetry analysis for the propagation vector **k** = (1/2 0 0) and the space group $P6_322$ gives four nonzero irreducible representations (Γs) for both the magnetic sites Co1(2$b$) and Co2(2$d$). All the IRs are one dimensional. The magnetic representation $\Gamma_{mag}$ for both the magnetic sites is composed of four IRs as

$$\Gamma_{Mag}^{Co(1),Co(2)} = \Gamma_1^1 + \Gamma_2^1 + \Gamma_3^1 + \Gamma_4^1 \tag{1}$$

The IRs $\Gamma_1$ and $\Gamma_4$ appear only once, whereas, $\Gamma_2$ and $\Gamma_3$ are repeated two times in the magnetic representation. The basis vectors of these IRs (the Fourier components of the magnetization) for two sites Co(1) and Co(2) are given in Table IV. The basis vectors are calculated using the projection operator technique implemented in the BASIREPS program [45]. Each of $\Gamma_1$ and $\Gamma_4$ has one basis vector, whereas, $\Gamma_2$ and $\Gamma_3$ have two basis vectors [Table IV].

Out of the above four IRs, the best refinement of the magnetic diffraction pattern is obtained for $\Gamma_2$. The refinable parameters are reduced to the coefficients of the basis vectors. It should also be noted that a *hkl*-dependent peak broadening of the magnetic Bragg peaks, especially for the (0,0,*l*)+**k** peaks, is found. The *hkl*-dependent widths of the magnetic peaks were simulated using the size formalism present in the FULLPROF suite.

The size effect (simulating the direction dependence of the magnetic correlations) was fitted using the size model 19 of the FULLPROF suite. Furthermore, the values of line widths for the first three strong magnetic reflections were given separately (refinable parameters) in order to reduce their influence on the rest of the peak shape refinement. The refined pure magnetic diffraction pattern as well as the total diffraction pattern at 1.8 K are shown in Fig. 5(c) and Fig. 5(b), respectively. For the refinement of the pure magnetic scattering the scale factor and the peak shape parameters were fixed to the values as obtained from the refinement of the purely nuclear data at 30 K.

    The corresponding magnetic structure is shown in Fig. 6. The magnetic moments are pointing along the crystallographic *b* direction. The moments are arranged to form collinear zigzag ferromagnetic chains along the *b* axis within the *ab* plane. Such zigzag FM chains are arranged antiferromagnetically perpendicular to the *b* axis. In this magnetic structure, out of six spins within a honeycomb unit three consecutive spins are arranged along one direction and the other three consecutive spins are arranged opposite to the first three spins. Therefore, for a given spin, out of three nearest neighbors two spins are parallel and one spin is antiparallel. Such antiferromagnetic honeycomb layers are coupled antiferromagnetically along the *c* axis. The observed zigzag AFM structure of honeycomb lattice cannot be explained by the sole existence of a $J_1$ as for this case the ground state should be a Néel type antiferromagnet having all antiparallel nearest neighbor spins. The collinear zigzag AFM state in a honeycomb lattice is a result of "order-by-disorder" phenomenon as outlined in the Introduction section. As predicted by several theoretical studies, the zigzag AFM ground state in a honeycomb lattice is possible in the presence of competing interactions $J_1$, $J_2$ and $J_3$ [1, 4]. Therefore, the zigzag magnetic ground state of the studied compound $Na_2Co_2TeO_6$ indicates the presence of NN, NNN, and NNNN interactions. Such zigzag AFM ground state has been experimentally reported recently for other frustrated honeycomb compounds $Cu_3Co_2SbO_6$ [16] with $S = 3/2$, and $Cu_3Ni_2SbO_6$ [16] with $S = 1$, $\alpha$-$RuCl_3$ [37, 47] and $Na_2IrO_3$ [48] with $j_{eff} = 1/2$. An experimental study of the spin-wave excitations of the $Na_2IrO_3$ compound having the zigzag AFM ground state showed that

substantial exchange couplings up to NNNN are required to explain the observed dispersion [49].

The refined ordered moment values were obtained to be $m_{Co(1)}$ = 2.77(3) $\mu_B$/Co$^{2+}$, and $m_{Co(2)}$ = -2.45(2) $\mu_B$/Co$^{2+}$ at 1.8 K. The ordered moment values are found to be significantly reduced from the theoretically expected spin only ordered moment value of 3 $\mu_B$ for Co$^{2+}$ ions (3$d^7$, $S$=3/2). It may be noted that in a diffraction measurement only the static components of the magnetic moments are detected, and the reduced moment can be a result of spin fluctuations that originate from spin frustrations. Reduced order moment values were also reported for other honeycomb compounds having zigzag AFM structures, i.e, 1.9(2) $\mu_B$/Ni$^{2+}$ ($S$ = 1) for Cu$_3$Ni$_2$SbO$_6$ [16], 2.4(1) $\mu_B$/Co$^{2+}$ ($S$ = 3/2) for Cu$_3$Co$_2$SbO$_6$ [16], and 0.5-0.6(1) $\mu_B$/Ru$^{3+}$ ($j_{eff}$ = 1/2) for $\alpha$-RuCl$_3$ [37, 47]. The reduction of the ordered moments was referred to the presence of spin fluctuations as well as of structural stacking faults of the honeycomb layers. In the present compound, the presence of a structural stacking fault is not evident; however, the presence of disorder is clear (discussed below in the next paragraph). The temperature dependent ordered magnetic moments for both the magnetic sites Co(1) and Co(2) are shown in the inset of Fig. 5 (c). Both sublattices order at the same temperature $T_N$ ~ 24.8 K, and the ordered moment values increase sharply below $T_N$ and saturate below ~ 10 K.

The temperature dependences of the two low-$Q$ magnetic peaks (0,0,0)+**k** and (0,0,1)+**k** are shown in Fig. 7. It becomes apparent that the second peak (0,0,1)+**k** (originated from both *ab* plane and *c*-axis out of plane magnetic correlations) is broader than the first peak (0,0,0)+**k** (arises solely due to the magnetic correlations within the ab plane) for the whole temperature range down to 1.8 K. The estimated widths (FWHM) for these two peaks are plotted as a function of temperature in the inset of Fig. 7. With decreasing temperature, the width of the first peak (0,0,0)+**k** decreases sharply just below the $T_N$, and then becomes constant with a value ~ 0.018 Å$^{-1}$) with further lowering of temperature. On the other hand, the width of the second magnetic peak (0,0,1)+**k** is found to be larger (~ 0.035 Å$^{-1}$ at 24.5 K), and remains constant with decreasing temperature. The whole set of peaks with (0,0,$l$)+**k** indices is also found to be broadened over the full temperature range. At the same time, nuclear Bragg peaks are found to be sharp and instrumental resolution limited [Fig. 5], hence, a structural stacking fault of the honeycomb layers can be ruled out in the present compound. In the presence of structural stacking fault, anomalous broadening of nuclear Bragg peaks are expected, as reported for related layered honeycomb compounds $Li_3Ni_2BiO_6$ [25], $Cu_3Ni_2SbO_6$, and $Cu_3Co_2SbO_6$ [16]. The observed broadening of the set of (0,0,$l$)+**k** magnetic Bragg peaks happens when the magnetic correlation along the $c$ axis is limited. The limited/restricted magnetic correlations along the $c$ axis may occur in the studied compound $Na_2Co_2TeO_6$ due to the presence of disorder in the intermediate Na layers between the magnetic honeycomb layers. Each disorder disrupts the magnetic coupling between the magnetic ions from the adjacent honeycomb layers, and leads hence to a partial breaking of the magnetic correlation along the $c$ axis. The presence of disordered distribution of Na ions among the three sites in the present compound $Na_2Co_2TeO_6$ is evident and already discussed in the crystal structure section. All the three Na-sites are partially occupied, and these Na triangular prismatic sites connect two adjacent magnetic honeycomb layers differently (by either sharing faces or edges with $CoO_6$/$TeO_6$ octahedra) along the $c$ axis [Fig. 2 (d-f)]. Therefore, the disordered statistical distribution of Na ions among the three Na sites strongly affects the magnetic coupling between the honeycomb layers. The present understanding, therefore, implies that the broken magnetic correlations between the honeycomb layers may be present for all structurally related P2 type compounds having partially occupied Na/Li sites. On the other hand, such a phenomenon is expected to be absent in the related O3 type compounds having fully occupied Na/Li sites. However, unavailability of proper data in literature prevents us from making a direct comparison. Therefore, detailed neutron diffraction experiments on both P2 and O3 type compounds are called for.

### D. Short-range magnetic correlations

We now discuss the persistence of short-range magnetic correlations above $T_N \sim 24.8$ K in $Na_2Co_2TeO_6$ which became visible in the neutron diffraction patterns at 25, 27, 35, 50, and 75 K measured on the high intensity powder diffractometer D20 at ILL, Grenoble, France and shown in Fig. 8. With decreasing temperature broad diffuse magnetic peaks, corresponding to short-range spin-spin correlations, appear below ∼ 50 K; at a temperature almost twice the Néel temperature $T_N$. A sharp decrease of the $\chi T$ values is also found at $T \sim 50$ K [Fig. 3]. The broad peaks with maximum at $Q \sim 0.7$ and 1.9 Å$^{-1}$ are situated at the same $Q$ positions where most intensed magnetic Bragg peaks are found below the $T_N \sim 24.8$ K. This indicates that the magnetic periodicity in the short-range state above $T_N$ is similar to the one in the long-range state below $T_N$. With decreasing temperature, the broad peaks grow monotonically down to $T_N$, before transforming into magnetic sharp Bragg peaks below $T_N$.

Similar broad diffuse magnetic peaks in neutron diffraction patterns were reported for several quasi-2D layered spin systems [50–55]. The short-range magnetic correlations were assigned to either 2D or 3D type depending on the profiles of the diffuse peaks. In case of 2D correlations, the peak shape is an asymmetric saw-tooth type which can be defined by a Warren function [50–53]. On the other hand for 3D correlations, peaks are symmetric, and can be defined by a Lorentzian function [53, 54]. In the present case, an asymmetric type peak shape is evident. However, the peak shape is more complex than the simple Warren function. Moreover, due to presence of two closely spaced magnetic peaks [i.e., (0,0,0)+**k**) at ∼ 0.69 Å$^{-1}$, and (0,0,1)+**k** at ∼ 0.89 Å$^{-1}$ for the first diffuse peak] in the present patterns it is difficult to find the dimensionality of the magnetic ordering from the simple fittings of the analytical functions like Warren or/and Lorentzian functions.

To analyze the diffuse scattering data we have used the program SPINVERT [36] which was successfully applied recently to several frustrated magnetic systems showing diffuse magnetic scatterings [56–58]. This program uses a RMC algorithm to fit the experimental powder data (pure magnetic pattern) by a large configuration of spin vectors. A key point about the RMC method is that it is entirely independent of a spin Hamiltonian. Therefore, it is not necessary to assume a form of the Hamiltonian to model the spin correlations. At the same time, it has the limitation that it does not produce a microscopic spin model as an output like the Rietveld method. Compared to other model-independent techniques for the analysis of diffuse neutron scattering (such as simple curve fitting), the RMC approach is superior in both quantity and accuracy of information it provides. Most importantly, this method provides real space spin-spin correlations. Furthermore, the SPINVERT program also calculates scattering profiles in the selected reciprocal planes by using the fitted spin configuration and the crystal structural information. As the program SPINVERT works with orthogonal axes, we have converted the hexagonal unit cell to an equivalent orthorhombic cell having twice the number of magnetic atoms. The transformation matrix for this case is given by

$$\begin{bmatrix} a' \\ b' \\ c' \end{bmatrix} = \begin{bmatrix} 1 & 0 & 0 \\ 1 & 2 & 0 \\ 0 & 0 & 1 \end{bmatrix} \begin{bmatrix} a \\ b \\ c \end{bmatrix}$$

In the present calculations, a supercell of 10×10×8 (6400 spins) of the orthorhombic crystal structure is generated, and a randomly oriented magnetic moment is assigned to each magnetic Co sites. The positions of spins are fixed at their crystallographic sites throughout the refinement, while their orientations are refined in order to fit the experimental data. A total of 1000 moves per spin are considered for each of the calculations. Ten individual fittings have been performed for each temperature to ensure the robustness of the results.

The calculated diffuse magnetic scattering intensities are shown in Fig. 9(a-c) by the solid lines along with the experimental data (filled circles) at 25, 27, and 35 K. The resulting spin configurations were used to reconstruct the $Q$-dependence of the diffuse scattering in the ($hk0$), ($h0l$) and ($0kl$) scattering planes [Figs. 9(d-l)] by using the SPINDIFF program extension to the SPINVERT program. Rod-like diffuse scattering is evident along the (00$l$) direction for both the ($h0l$) and ($0kl$) scattering planes at all three temperatures. The rod-like feature becomes sharper with decreasing temperature, and gets confined around $h = \pm(2n+1)/2$ in the ($h0l$) plane, and $k = \pm(2n+1)$ in the ($0kl$) plane, where $n$ is an integer. The positions where the rod-like scattering is found are in agreement with the propagation vector **k** = (1/2 0 0) of the magnetic ordered state below $T_N$. The rod-like scattering reveals that the magnetic correlations are confined within the 2D honeycomb planes (*ab* plane). In this case no restriction is imposed on the *l* value, which leads to a rod-like scattering along *l*. Within the (*hk*0) plane, the symmetric type of scattering suggests an isotropic correlation within the honeycomb planes. The above results confirm the existence of a 2D magnetic ordering within the honeycomb layers (*ab* planes) at all temperatures 35, 27, and 25 K above the $T_N$. The 2D magnetic correlations, confined within the *ab* plane, are indeed expected from the layered type crystal structure of the present compound. The crystal structure provides stronger intraplane interactions via the Co-O-Co superexchange interaction pathways, and relatively weak interplane interactions via the Na-layers along the *c* axis. In addition, the disorder in the intermediate Na layers which interrupt the magnetic couplings between honeycomb layers also favours the 2D magnetic correlations within the *ab* planes.

For further understanding of the nature of the short-range magnetic ordering, the real space spin-pair correlation functions $\langle \vec{S}(0) \cdot \vec{S}(r) \rangle$ are calculated, and shown in Fig. 10. Each data point in Fig. 10 corresponds to a distance between two magnetic sites within the Na$_2$Co$_2$TeO$_6$ crystal structure. The spin-pair correlation functions are calculated from the fitted spin configurations by using the program SPINCORREL (an extension of the SPINVERT program). A larger absolute value of $\langle S(0).S(r) \rangle$ indicates a stronger preference for a collinear arrangement of the spins, separated by a distance *r*. The sign of $\langle S(0).S(r) \rangle$ indicates whether the spins are parallel (+) or antiparallel (-) to each other. The spin-pair correlations for the present compound (Fig. 10) include both positive and negative values that decrease with the increasing distance (*r*), and almost vanish at a distance of ~ 12 Å. This is an indication of short-range AFM correlations. The average NN spin-spin correlation (*r* = 3.0416 Å) is found to be FM which is in agreement with the ordered magnetic structure below $T_N$; where two out of three NN spin-pair correlations are FM (Fig. 6). Both the NNN (*r* = 5.2679 Å) and NNNN (*r* = 6.0827 Å) spin-pair correlations within the honeycomb plane are found to be AFM which is also consistent with the ordered magnetic structure. In this case, four out of six NNN spin-pair correlations are AFM; and all three NNNN spin-pair correlations are AFM (Fig. 6). The temperature dependences of these three correlations are shown in Fig. 10(b). With deceasing temperature, an increase of the correlations without any change in their signs is evident. The spin-pair correlation between the honeycomb planes along the $c$ axis ($r$ = 5.586 Å) is found to be AFM, which is again consistent with the ordered magnetic structure below the $T_N$. All the above results indicate a similar nature of magnetic symmetry in both the short-range (above the $T_N$) and long-range (below the $T_N$) ordered states. This indeed reveals that the evolution of magnetic correlations as a function of temperature in $Na_2Co_2TeO_6$ is governed by the intermediate Na layers.

Now, we discuss the consequence of having observed a zigzag AFM ground state in the present honeycomb lattice compound $Na_2Co_2TeO_6$. As mentioned earlier that such a zigzag magnetic ground state was reported for other honeycomb lattice compounds $\alpha$-RuCl$_3$ [37, 47] and $Na_2IrO_3$ [48] with an effective spin $J_{eff}$ = 1/2. These compounds were reported to situate proximately to the Kitaev spin liquid state [13, 59] where the combination of isotropic Heisenberg exchange interaction and anisotropic Kitaev term through strong spin-lattice coupling gives rise to exotic behaviors. Such a situation can be considered for the present compound with magnetic ions $Co^{2+}$ having the similar zigzag AFM ground state. According to the Hunds rules, for the free $Co^{2+}$ ($3d^7$) ion having 7 electrons the total orbital and spin angular momenta are $L$ = 3 and $S$ = 3/2, respectively. In a distorted (triangular) octahedral environment [as found for the present compound, and discussed in the Crystal structure section], the orbital and spin degrees of freedoms are entangled by spin-orbit coupling which makes the total angular momentum a conserved quantity. Here, the lowest-lying Kramers doublet of $Co^{2+}$ is well separated from the higher-lying spin-orbit quartet and sextet. Thus the magnetic moment of $Co^{2+}$ can be considered as an effective $J_{eff}$ = 1/2 pseudospin with a large anisotropy [60]. Thus $Na_2Co_2TeO_6$ could be an analogous to the above two compounds, hence, there is a possibility to realize the Heisenberg-Kitaev model in this compound as well. The additional advantage of the present compound is that the magnetic coupling between the honeycomb layers could easily be tuned by varying the Na concentrations.

The results of the present work provide an insight into the crystal and magnetic structural correlations in the layered honeycomb lattice compound $Na_2Co_2TeO_6$. One of the unique aspects of the present work is the understanding of the detailed microscopic magnetic correlations as a function of temperature both above and below the $T_N$. We have demonstrated that the crystallographically soft Na layers dictate the formation of magnetic correlations. This study provides experimental evidence for the theoretical ideas to explain the nature of the magnetic correlations in a honeycomb lattice, a fertile ground yet to be fully explored. The results of the present study are expected to open up future studies on isostructural compounds having varying spin values and magnetic interaction strengths.

## IV. SUMMARY AND CONCLUSIONS

We have investigated the structural and magnetic properties of the new frustrated layered honeycomb lattice compound $Na_2Co_2TeO_6$. Our low temperature neutron diffraction investigation reveals the existence of a zigzag AFM long-range ordered state below $T_N \sim 24.8$ K which has a restricted correlation along the $c$ axis even deep inside the ordered state at 1.8 K. The restricted correlation along the $c$ axis occurs due to broken magnetic connections, inducted by the disorder distribution of Na ions between three partially occupied sites within the interconnecting layers. Here, Na ions form $NaO_6$ triangular prism in all these three sites, and connect two adjacent magnetic honeycomb layers differently either by sharing faces or edges with $CoO_6/TeO_6$ octahedra along the $c$ axis. Reduced ordered moments of $m_{Co(1)}$ = 2.77(3) $\mu_B/Co^{2+}$ and $m_{Co(2)}$ = 2.45(2) $\mu_B/Co^{2+}$ are found at 1.8 K suggesting the persistent spin fluctuations in the ordered state. Our study also shows the presence of short-range magnetic correlations above $T_N$. The RMC analysis reveals that the dominant spin-pair correlations are within the honeycomb layers ($ab$-plane) with a correlation length about 12 Å at 25 K. The symmetry of the magnetic order is found to be identical in both the short-range and long-range ordered states. This study, thus, provides a comprehensive picture of the microscopic magnetic correlations over the entire temperature range covering the regions both above and below $T_N$. The present study also demonstrates that the magnetic correlations in $Na_2Co_2TeO_6$ are dictated by the intermediate nonmagnetic Na layers and provides an in-depth understanding of the crystal and magnetic structural correlations.

TABLE I. The Rietveld refined lattice constants, fractional atomic coordinates, and isotropic thermal parameters ($B_{iso}$) for $Na_2Co_2TeO_6$ at room temperature. Lattice constants $a$ = 5.2770(2) Å, $c$ = 11.2231(1) Å. *Occ.* stands for site occupancy.

| Atom  | Site | x/a       | y/b        | z/c       | Biso     | Occ.      |
|-------|------|-----------|------------|-----------|----------|-----------|
| Co(1) | 2b   | 0         | 0          | 0.25      | 0.32(6)  | 1.0       |
| Co(2) | 2d   | 2/3       | 1/3        | 0.25      | 0.68(10) | 1.0       |
| Te    | 2c   | 1/3       | 2/3        | 0.25      | 0.68(10) | 1.0       |
| O     | 12i  | 0.6446(5) | -0.0260(4) | 0.3438(2) | 0.98(2)  | 1.0       |
| Na(1) | 12i  | 0.698(3)  | 0.056(2)   | 0.003(2)  | 1.13(6)  | 0.225(2)  |
| Na(2) | 12i  | 0.361(9)  | 0.620(8)   | -0.024(2) | 1.13(6)  | 0.085(3)  |
| Na(3) | 2a   | 0         | 0          | 0         | 1.13(6)  | 0.153(3)  |

TABLE II. Possible pathways for NN, NNN, and NNNN exchange interactions $J_1$ and $J_2$ and $J_3$, respectively. The Co...Co direct distances, metal oxide ($M$–O) bond lengths and bond-angles for the exchange interactions $J_1$, $J_2$ and $J_3$ in Na$_2$Co$_2$TeO$_6$ at room temperature.

| Exchange interaction | Pathways | Co...Co direct distance (Å) | Bond lengths (Å) | Bond angles (deg.) |
|---|---|---|---|---|
| $J_1$ | Co(1)–O–Co(2) | Co(1)–Co(2) = 3.044(5) | Co(1)–O= 2.094(1) | Co(1)–O–Co(2)= 92.5(1) |
| | | | Co(2)–O= 2.119(1) | |
| $J_2$ | Co(1)–O–Co(2)–O–Co(1) /Co(1)–O–Te–O–Co(1) /Co(1)–O–O–Co(1) | Co(1)–Co(1)= 5.272(1) | Co(1)–O= 2.094(1) | Co(1)–O–Co(2)= 92.5(1) |
| | | | Co(2)–O= 2.119(1) | O–Co(2)–O= 174.5(2)/97.4(2)/78.5(1) |
| | | | Te–O= 1.941(1) | Co(1)–O–Te= 97.9(1) |
| | | | O–O= 2.683(2) | O–Te–O= 179.0(1) |
| | Co(2)–O–Co(1)–O–Co(2) /Co(2)–O–Te–O–Co(2) /Co(2)–O–O–Co(2) | Co(2)–Co(2) = 5.272(1) | Co(2)–O= 2.119(1) | Co(2)–O–Co(1)= 92.5(1) |
| | | | Co(1)–O= 2.094(1) | O–Co(1)–O=173.5(1)/96.9(1)/78.4(1) |
| | | | Te–O= 1.941(1) | Co(2)–O–Te= 97.0(1) |
| | | | O–O= 2.683(2) | O–Te–O = 179.0(1) |
| $J_3$ | Co(1)–O–Te–O–Co(2) | Co(1)–Co(2) = 6.088(5) | Co(1)–O= 2.094(1) | Co(1)–O–Te=97.9(1) |
| | | | Co(2)–O= 2.119(1) | Co(2)–O–Te= 97.0(1) |
| | | | Te–O= 1.941(1) | O–Te–O = 179.0(1) |

TABLE III. The local crystal structural parameters [bond lengths (Co–O) and bond angles (O– Co–O) within the octahedra] for the two cobalt sites, [Co(1) and Co(2)].

|  |  | Co(1) | Co(2) |
|---|---|---|---|
| bond length (Å) | (Co–O) | 2.094(1) | 2.119(1) |
| bond angle (°) | (O–Co–O) | | |
|  | diagonal | 173.5(1) | 174.5(2) |
|  | orthogonal | 78.4(1) | 78.5(1) |
|  |  | 88.1(1) | 86.8(1) |
|  |  | 96.9(1) | 97.4(2) |

TABLE IV. Basis vectors of the magnetic sites Co(1) and Co(2) with the propagation vector **k** = (1/2 0 0) for $Na_2Co_2TeO_6$. Only the real components of the basis vectors are presented. The four atoms of the nonprimitive basis are defined according to Co(1)-1:(0, 0, 0.25); Co(1)-2: (0, 0, 0.75); and Co(2)-1: (0.6667, 0.3333, 0.2500); Co(2)-2: (-0.6667, -0.3333, 0.7500).

| IRs | | Basis Vectors | | | |
| --- | --- | --- | --- | --- | --- |
| | | Site (2$b$) | | Site (2$d$) | |
| | | Co(1)-1 | Co(1)-2 | Co(2)-1 | Co(2)-2 |
| $\Gamma 1_1$ | $\Psi_1$ | (210) | (-2-10) | (210) | (-2-10) |
| $\Gamma 1_2$ | $\Psi_1$ | (0-10) | (010) | (0-10) | (010) |
| | $\Psi_2$ | (001) | (001) | (001) | (001) |
| $\Gamma 1_3$ | $\Psi_1$ | (0-10) | (0-10) | (0-10) | (0-10) |
| | $\Psi_2$ | (001) | (00-1) | (001) | (00-1) |
| $\Gamma 1_4$ | $\Psi_1$ | (210) | (210) | (210) | (210) |

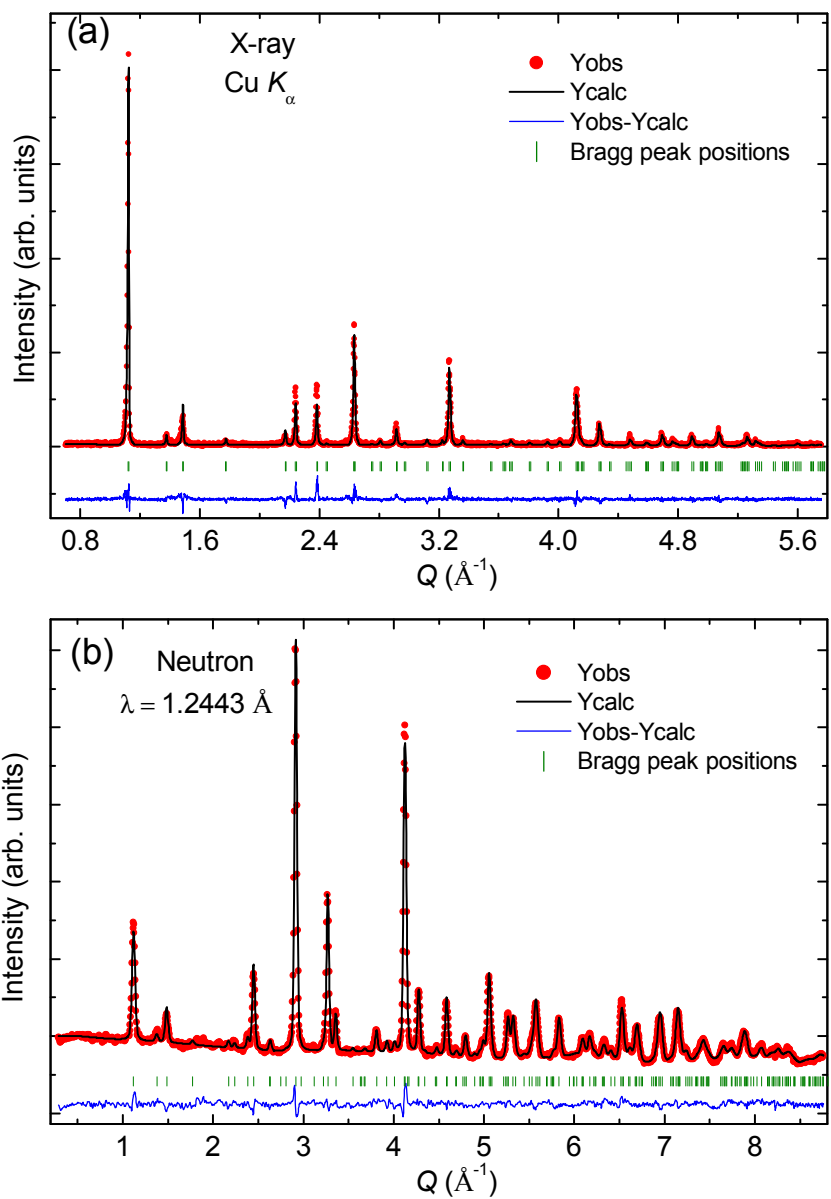

FIG. 1. (Color online) Experimentally observed (circles) and calculated (solid line through the data points) (a) x-ray and (b) neutron diffraction patterns (intensity vs momentum transfer $Q$) for $Na_2Co_2TeO_6$ at room temperature. The difference between observed and calculated patterns is shown by the solid lines at the bottom of each panel. The vertical bars indicate the positions of allowed nuclear Bragg peaks.

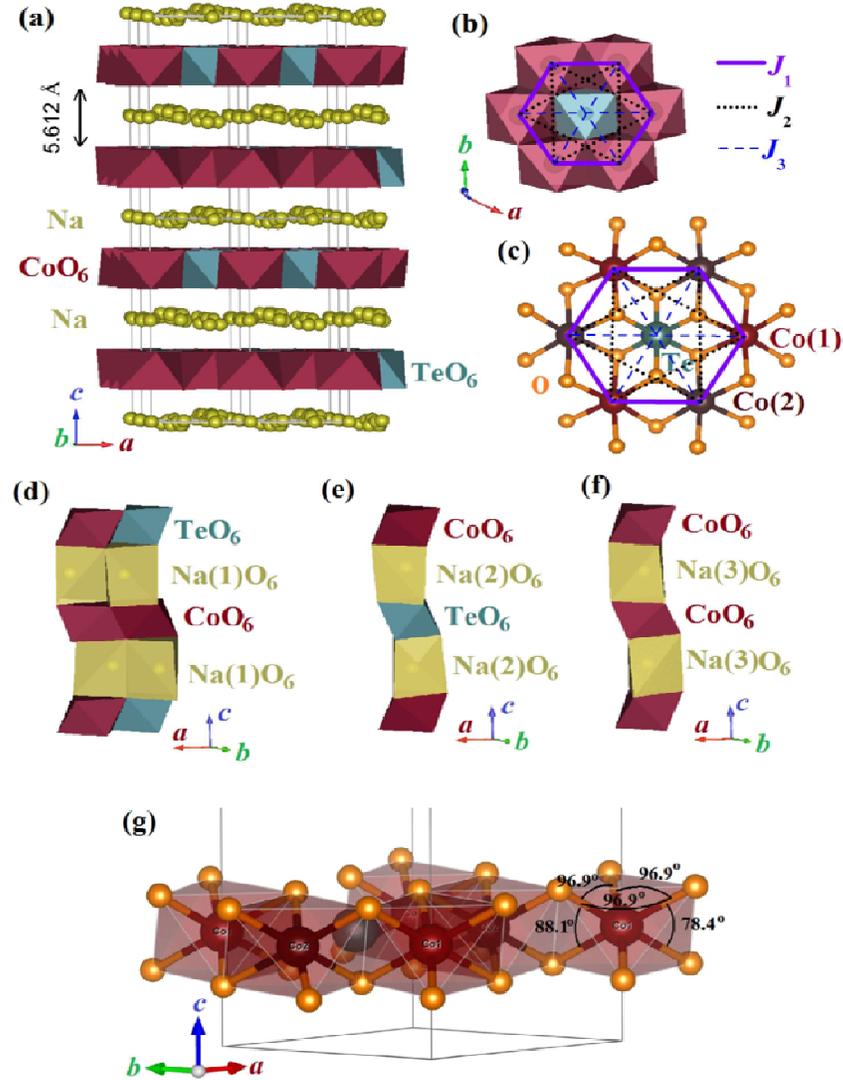

FIG. 2. (Color online) The crystal structure of $Na_2Co_2TeO_6$. (a) Stacking of the honeycomb layers along the $c$ axis. (b) The honeycomb lattices of Co with NN ($J_1$), NNN ($J_2$), and NNNN ($J_3$) interactions. The $TeO_6$ octahedra sit at the center of each honeycomb unit. (c) The local crystal structure showing a honeycomb unit with the atomic bonds. (d)-(f) The connection of the magnetic (honeycomb) layers along the $c$ axis by Na(1), Na(2) and Na(3) triangular prisms, respectively. (g) The local crystal structure within the basal plane showing the octahedral environment around the magnetic Co sites. The compression of $CoO_6$ octahedra along the $c$ axis leads to the decrease of O-Co-O bond angles across the shared edges within the layers.

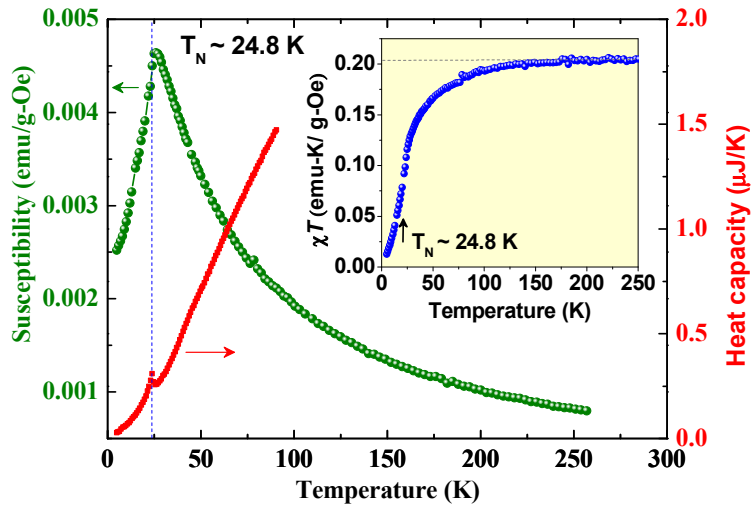

FIG. 3. (Color online) The temperature dependent ac susceptibility ($\chi$) and heat capacity ($C_p$) for Na$_2$Co$_2$TeO$_6$.

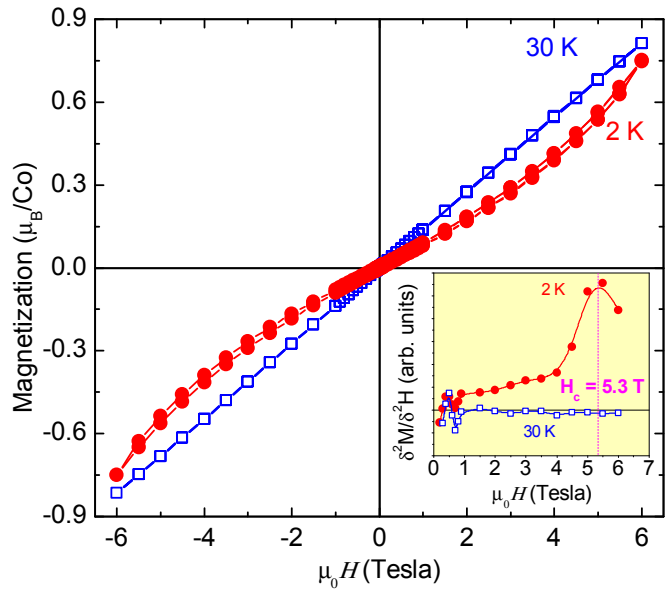

FIG. 4. (Color online) The isothermal magnetization of Na$_2$Co$_2$TeO$_6$ as a function of magnetic field at 2 and 30 K. The inset shows the second derivative of magnetization $\delta^2 M/\delta^2 H$ vs $H$ curves.

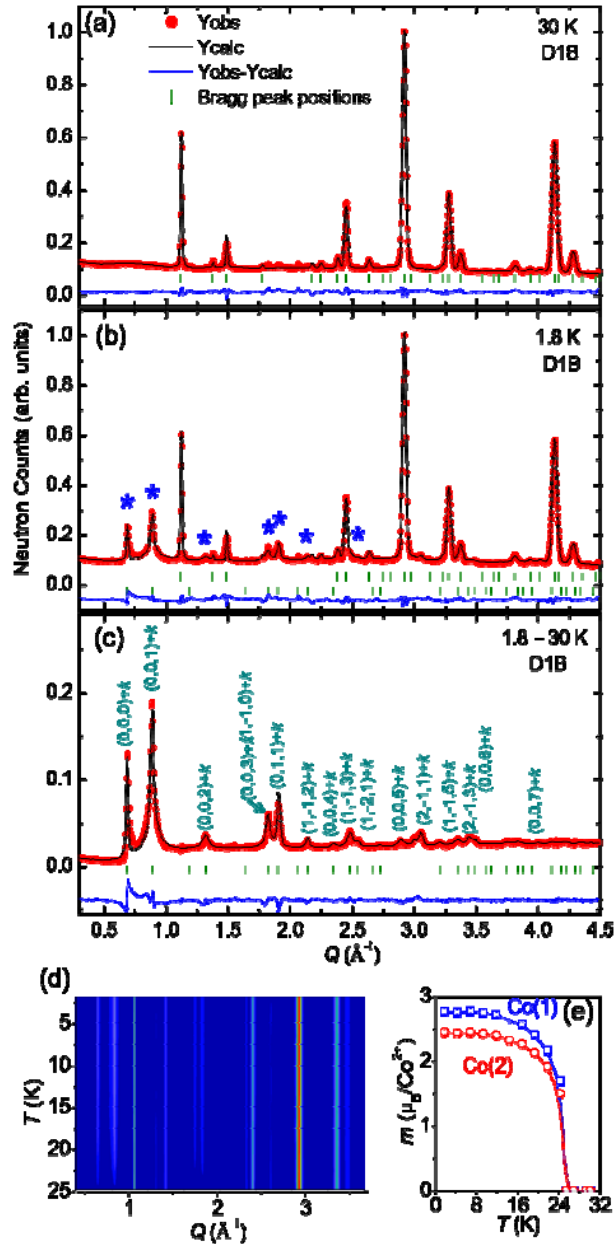

FIG. 5. (Color online) Experimentally observed (circles) and calculated (solid line through the data points) neutron diffraction patterns ( λ = 2.524 Å) for $Na_2Co_2TeO_6$ at (a) 30 K and (b) 1.8 K. (c) The pure magnetic pattern at 1.8 K after subtraction of nuclear pattern at 30 K. The difference between observed and calculated patterns is shown by the solid line at the bottom of each panel. The vertical bars indicate the positions of allowed nuclear and magnetic Bragg peaks, respectively. (d) The 2D color plot of the temperature dependent neutron diffraction patterns, showing the appearance of magnetic Bragg peaks below $T_N \sim$ 24.8 K. (e) The temperature dependent ordered magnetic moments for the Co(1) and Co(2) sites. Lines are guide to the eyes.

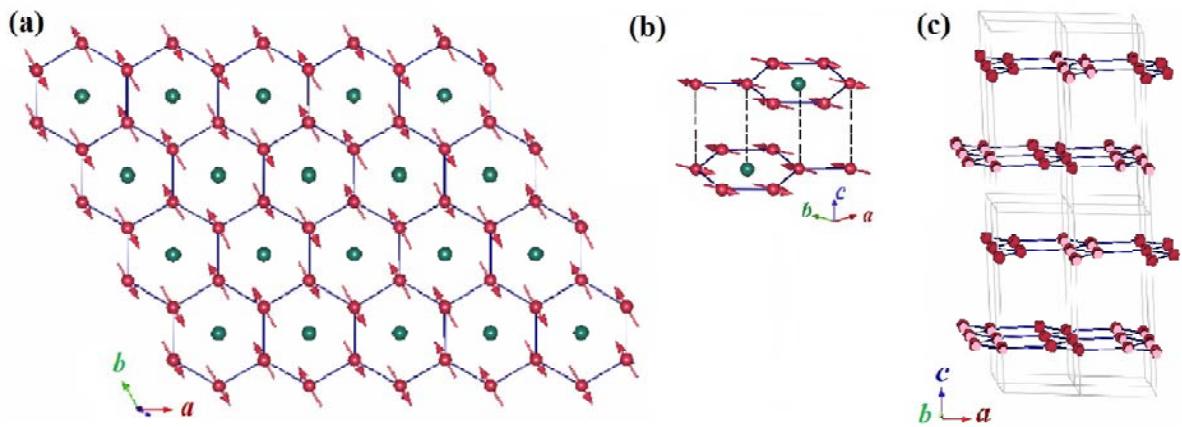

FIG. 6. (Color online) The magnetic structure of $Na_2Co_2TeO_6$. (a) The zigzag AFM spin arrangements of Co ions (red spheres) within the honeycomb lattice. Te atoms (green spheres) sit at the center of each honeycomb units. (b) The coupling of the two honeycomb planes along the $c$ axis within a unit cell. (c) The projection of the magnetic structure in the $ac$ plane. The dark and light colored circles represent the moment directions along the +ve $b$ axis and -ve $b$ axis, respectively. The thin gray lines show the dimension of the unit cells. Te atoms are omitted for clarity.

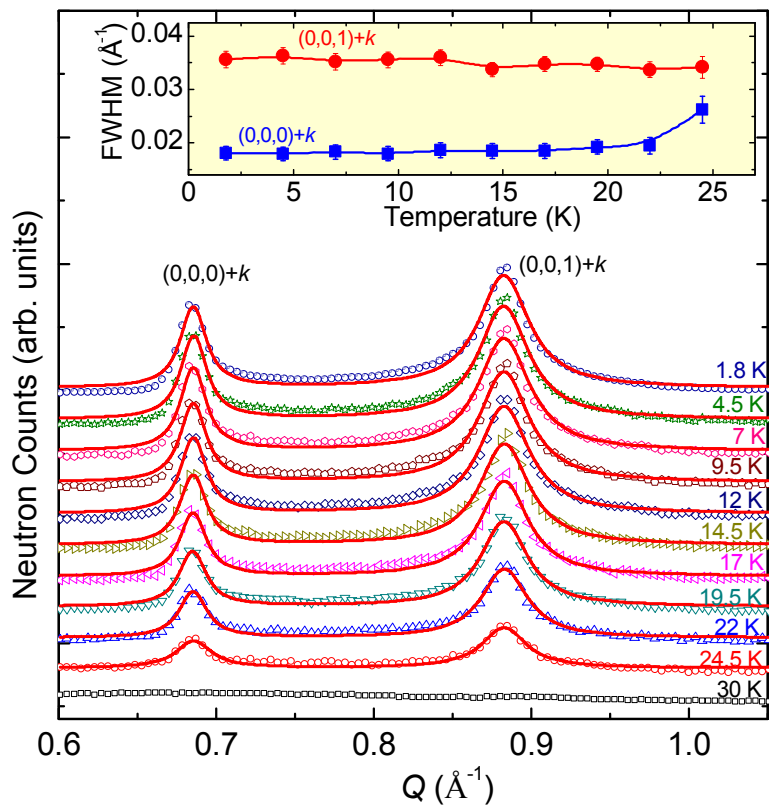

FIG. 7. (Color online) (a) The temperature evolution of the two low-$Q$ magnetic peaks $(0,0,0)+\mathbf{k}$ and $(0,0,1)+\mathbf{k}$. The solid lines are the fitted curves by a Lorentzian function. (b) The temperature dependent peak width (FWHM) of the two magnetic peaks.

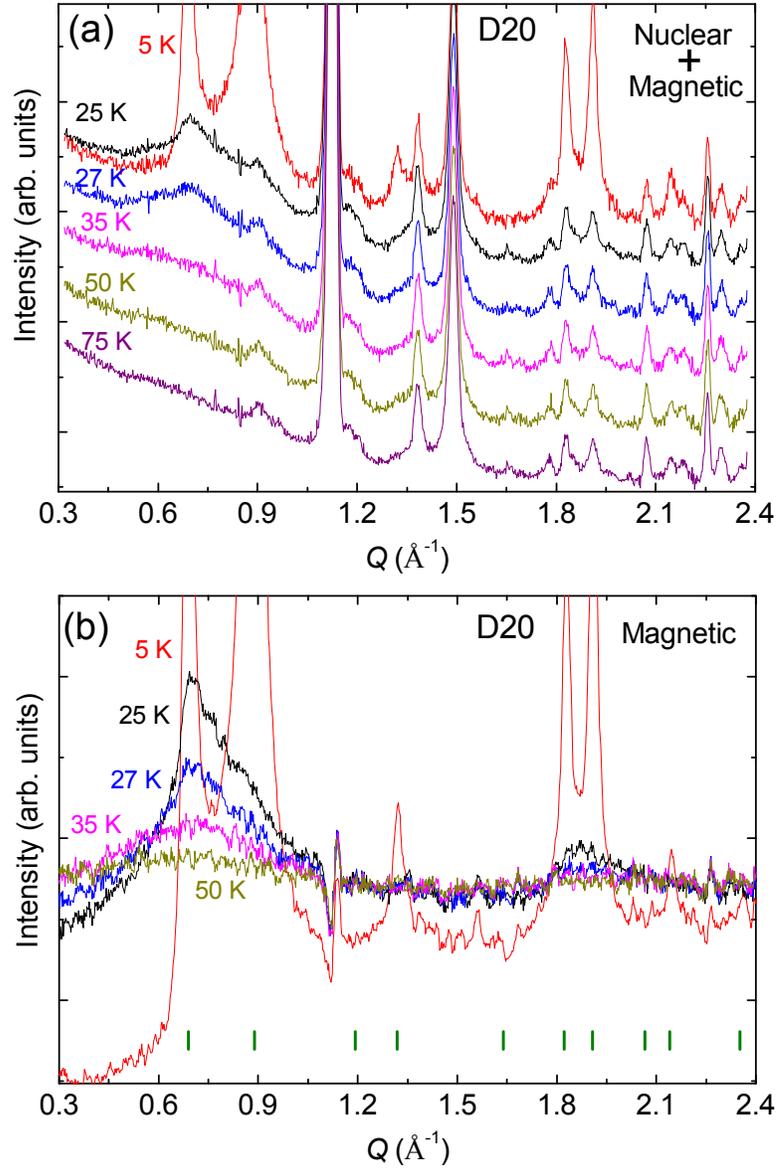

FIG. 8. (Color online) (a) The experimentally measured neutron diffraction patterns (λ = 2.41 Å) at 5, 25, 27, 35, 50, and 75 K. (b) The pure magnetic diffraction patterns at 5, 25, 27, 35, and 50 K after subtraction of the 75 K pattern as paramagnetic background. The vertical bars show the magnetic Bragg peak positions in the 3D long-range ordered state below $T_N \sim$ 24.8 K. The lower background in the (5 K - 75 K) pattern appears from the subtraction of the $Q$-dependent paramagnetic scatterings at 75 K arising due to the magnetic form factor of $Co^{2+}$ ions ($S$= 3/2).

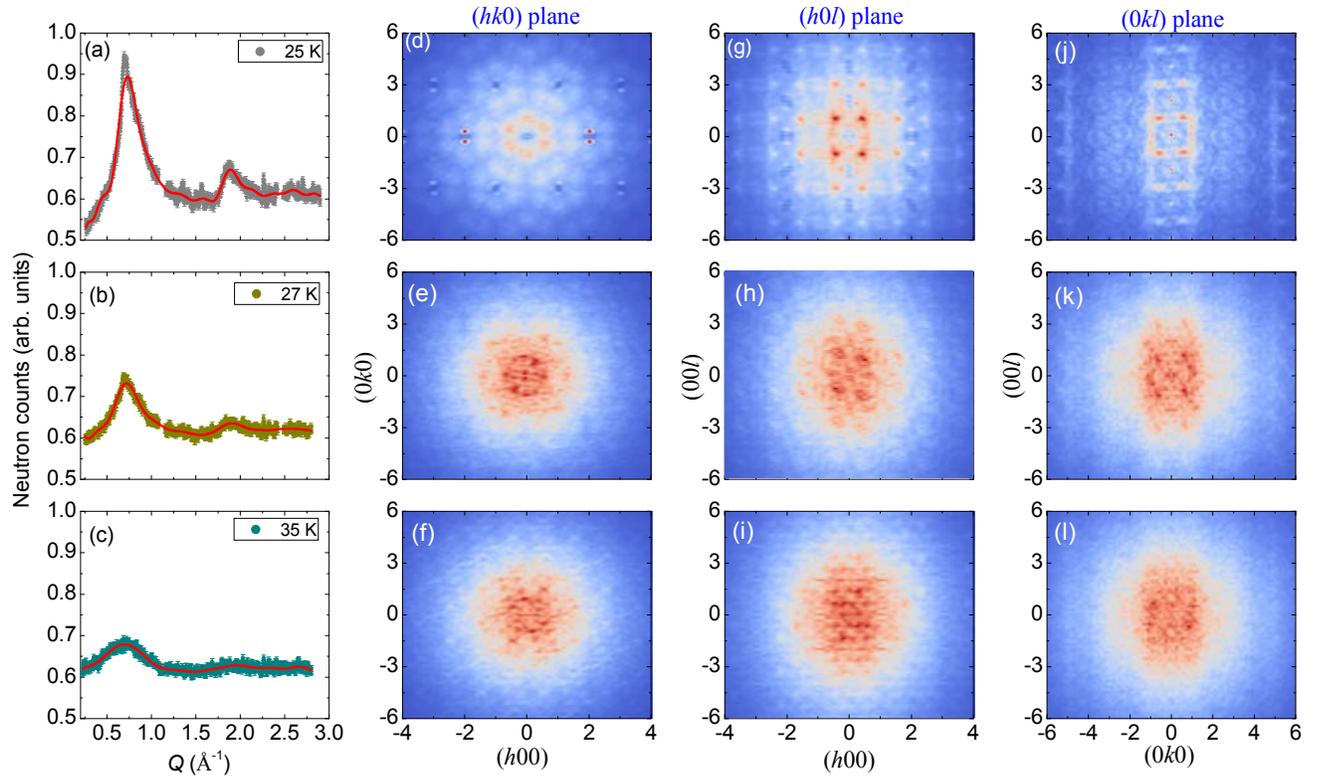

FIG. 9. (Color online) (a)-(c) The experimentally measured diffuse magnetic scattering at 25, 27, and 35 K after subtraction of the paramagnetic background at 75 K. The solid lines in each panel are the calculated scattering intensities by the RMC method. (d-l) The reconstructed diffraction patterns in the ($hk$0), ($h$0$l$) and (0$kl$) scattering planes.

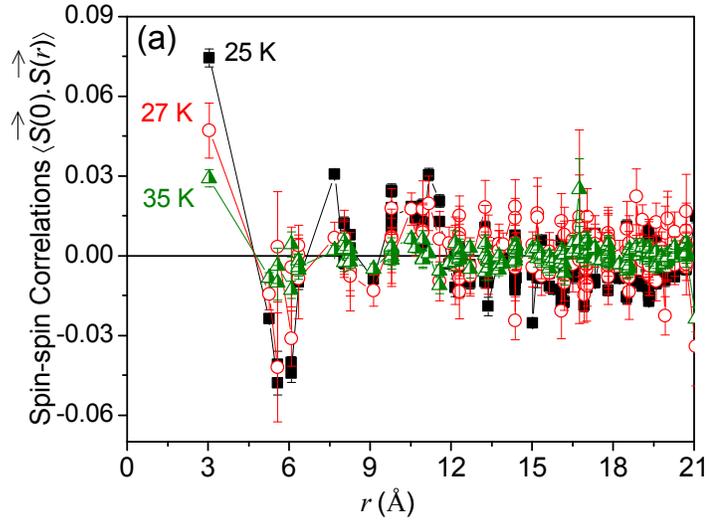

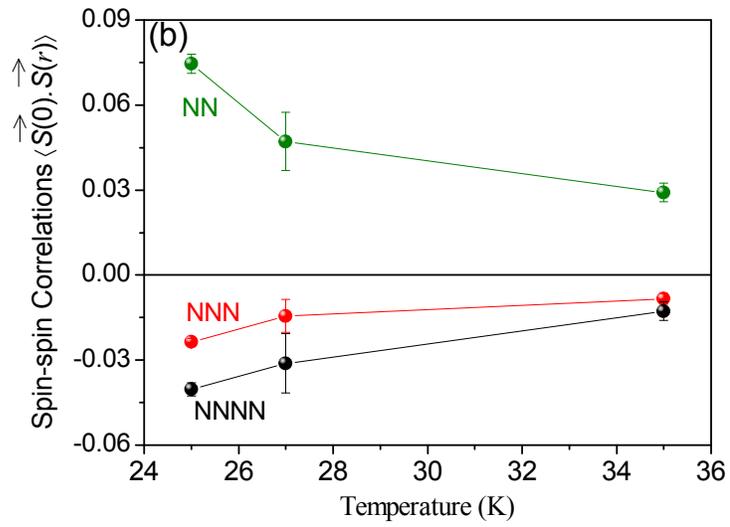

FIG. 10. (Color online) (a) The radial spin-pair correlation functions at 25, 27, and 35 K corresponding to the fits shown in Figs. 9 (a-c). (b) The temperature dependent spin-pair correlation functions for the NN, NNN, and NNNN within the honeycomb lattices (in the *ab* plane).